\documentclass[12pt,a4paper]{iopart}
\usepackage{graphicx,cite}
\eqnobysec

\begin{document}

\title[Resonance in Bose-Einstein condensate]{Resonance in
Bose-Einstein
condensate oscillation from a
periodic variation in scattering length}

\author{Sadhan K. Adhikari}
 \address{Instituto de F\'{\i}sica Te\'orica, Universidade Estadual
Paulista,
01.405-900 S\~ao Paulo, S\~ao Paulo, Brazil}

\begin{abstract}

Using the explicit numerical solution of the axially-symmetric
Gross-Pitaevskii equation we study the oscillation of the Bose-Einstein
condensate induced by a periodic variation in the atomic scattering length
$a$. When the frequency of oscillation of $a$ is an even
multiple of the radial or axial trap frequency, respectively, the radial
or axial
oscillation of the condensate exhibits resonance with novel feature.
In
this nonlinear problem without damping, at resonance in the steady
state the
amplitude of oscillation passes through maximum and minimum.  Such growth
and decay cycle of
the amplitude may keep on repeating. Similar behavior is also observed in
a rotating Bose-Einstein condensate.

\end{abstract}


\submitto{\jpb}

\maketitle

\section{Introduction}
 
The successful detection \cite{1,ex2} of Bose-Einstein condensates
(BEC) in
dilute
weakly-interacting bosonic atoms employing magnetic trap at ultra-low
temperature stimulated intense theoretical studies on different aspects of
the condensate \cite{11,11a,11b}. The experimental magnetic trap could be
either
spherically symmetric or axially symmetric.  The properties of an ideal
condensate at zero temperature are usually described by the
time-dependent, nonlinear, mean-field Gross-Pitaevskii (GP) equation
\cite{2} which incorporates appropriately the trap potential as well as
the interaction between the atoms forming the condensate.

One of the problems of interest in Bose-Einstein condensation  is the
study of the oscillation in the condensate induced by a sudden change in
the interaction acting on it. There have been both experimental \cite{5} 
and
theoretical \cite{6} studies of the motion of the condensate under a
sudden or
continuous variation of the trapping frequencies. There has also been
similar studies induced by a sudden variation of the atomic scattering
length. In the latter class there have been experimental studies of
explosive emission of atoms from a condensate when the scattering length
is suddenly turned negative (attractive) from positive (repulsive), thus
leading to collapse and explosion of the condensate \cite{7}. This could
also lead
to the observation of a soliton train \cite{8}.
There have been
theoretical attempts to explain the dynamics of the condensate 
for collapse and explosion \cite{9} as well as for soliton train \cite{9a}
using the GP equation.

Resonance is an interesting feature of any oscillation under the action of
an external periodic force manifesting in  a  very
large amplitude. Usually, a resonance appears when the frequency of the
external
periodic force becomes equal to that of the natural oscillation of the
system \cite{symon}. Such a resonance appears in many areas of
physics. There have been
studies of resonance in different problems of linear dynamics and the
appearance of resonance is well understood in these problems both
analytically and numerically. The investigation of resonance in nonlinear
dynamics  is far more interesting and nontrivial. Consequently, 
resonances in nonlinear problems  are not so well
understood. This makes
the study of resonance in the dynamics generated by the nonlinear GP
equation of general interest. Apart from mere theoretical interest, such
resonance in the oscillation of the BEC governed by the GP equation can
also be studied experimentally so that one can compare theoretical
prediction with experiment.

The possibility of generating a resonance in the oscillation of a BEC
subject to a varying trapping potential has been investigated
theoretically \cite{6}. The variation of trapping potential in the GP
equation
corresponds to a linear external perturbation. Here we investigate the
resonance dynamics of the BEC subject to a periodic variation of the
scattering length, which should be considered as a nonlinear 
periodic force acting in the GP equation. Such a variation of the
scattering length is possible
near a Feshbach resonance by varying an external magnetic field
\cite{fesh}. For experimentalists  a periodic 
  variation    of   the external   magnetic   field is easier to
implement and the magnetic field is nonlinearly related to the scattering
length. However,  a simple periodic  variation of the
  scattering  length  is  a  cleaner  theoretical  possibility  and we
shall apply
such a variation in this study.
 The appearance of a
resonance in the oscillation of a BEC due to a
periodic variation of the
scattering length 
has been postulated recently in a spherically symmetric trap
\cite{abd}. Here we present a  complete study of the resonance
dynamics in this case. In
addition, we extend this
investigation to the more realistic and complicated case of an axially
symmetric trap as well as to vortex states in such a  trap.

In the simplest case of a damped classical oscillator under a periodic
external force with the natural frequency of oscillation, the resonance
appears due a constant phase difference between the oscillation and the
external force \cite{symon}. This leads to a gradual increase in kinetic
energy over
each cycle of oscillation during an interval of time and hence in the
amplitude of oscillation in this interval.  
Eventually, the rate of loss of energy due to damping is exactly
compensated for by the increase in energy from the external force and the
oscillator oscillates with a large constant amplitude at large times at
resonance. The constant phase difference between the external force and
the oscillation is possible only at resonance when the two frequencies
match and is necessary for resonance.

The situation is much more complicated in the case of a trapped BEC
subject to a nonlinear external force due to a periodic variation in the
scattering length. When subject to a sudden perturbation, the natural
frequency of oscillation of a trapped spherical BEC governed by the GP
equation with nearly zero nonlinearity is $2\omega$ where $\omega$ is the
frequency of the existing trap \cite{7,ska}. This result is  
exact in the
linear case $n=0$. However, for the  nonlinear problem $(n\ne 0)$, this
natural frequency may change. In this paper we shall only be considering small
nonlinearity where the natural frequencies can be taken approximately as
$2\omega$. 
 In physical analogy with the
classical oscillator, a resonance is expected in the BEC oscillation when
the scattering length oscillates with frequency $2\omega$ or its
multiples. In our study we find that this is indeed the case. However, a
much richer dynamics emerges in this case compared to the case of linear
systems. We find that due to nonlinearity, the phase difference $\delta$
between the BEC oscillation and the external nonlinear force varies with
time. Consequently, during a certain interval of time the BEC may steadily
gain energy from the external force in all cycles and the amplitude of
oscillation may grow leading to a resonance. In the case of BEC there is
no
damping and for a constant phase difference $\delta$ the amplitude is
expected to grow beyond limit. However, as a result of the variation of
the phase $\delta$ with time due the nonlinearity, after some time the
system starts to lose energy due to the action of the external force.
Eventually, the oscillation of the the system can virtually stop. By that
time the modified phase difference favors the increase of energy and the
amplitude starts to grow again. Consequently, the resonance of the system
has a peculiar nature; the amplitude of oscillation passes through
pronounced maxima and minima. In a single classical damped oscillator this
is not possible under the action of a periodic force \cite{symon}.
However, this is possible in case of a coupled system with exchange of
energy between the two oscillators.

We also investigate the resonant oscillation of an axially symmetric BEC
due to a periodic variation of the scattering length. When the frequency
of oscillation of the scattering length equals the natural frequency of
the radial or axial oscillation of the BEC, we find resonance  
in the radial or axial oscillation, respectively.

In section  2 we describe briefly the time-dependent  GP equation with 
spherical and axial traps. The essential details of the 
split-step Crank-Nicholson method used for numerical solution   is also 
described in this section.  In section 3  we report the
numerical results of the present investigation of resonance 
for the spherically 
symmetric case. In section 4 we present the same for the
axially symmetric case for condensates with zero angular momentum as well
as for vortex states with nonzero angular momentum.
 Finally, in  section 5 we present 
a discussion and summary of our study.

\section{Nonlinear Gross-Pitaevskii Equation}

At zero temperature, the time-dependent Bose-Einstein condensate wave
function $\Psi({\bf r};\tau)$ at position ${\bf r}$ and time $\tau $ may
be described by the following  mean-field nonlinear GP equation
\cite{11,8}
\begin{eqnarray}\label{a} 
\biggr[ -\frac{\hbar^2\nabla ^2}{2m}
+ V({\bf r})  
+ gN|\Psi({\bf
r};\tau)|^2
-i\hbar\frac{\partial
}{\partial \tau} \biggr]\Psi({\bf r};\tau)=0.   \end{eqnarray} Here $m$
is
the mass and  $N$ the number of atoms in the
condensate, 
 $g=4\pi \hbar^2 a/m $ the strength of interatomic interaction, with
$a$ the atomic scattering length. The normalization condition of the wave
function is
$ \int d{\bf r} |\Psi({\bf r};\tau)|^2 = 1. $

\subsection{Spherically Symmetric Case}

In this case  the trap potential is given by $  V({\bf
r}) =\frac{1}{2}m \omega ^2r^2$, where $\omega$ is the angular frequency
and $r$ the radial distance. The wave function can be written as
$\Psi({\bf r};\tau)=\psi(r,\tau)$. 
 After a transformation of variables to 
dimensionless quantities 
defined by $x =\sqrt 2 r/l$,     $t=\tau \omega, $
$l\equiv \sqrt {(\hbar/m\omega)} $ and
$ \varphi(x,t) =x\psi(r,\tau)(4 
\pi l^3/\sqrt 8)^{1/2}$, the GP equation in this case becomes
\begin{eqnarray} \label{c}
\left[-\frac{\partial^2}
{\partial x^2}+\frac{x^2}{4}+2\sqrt 2 {n}\left|
\frac{\varphi(x,t)}{x}
\right| ^2 -i\frac{\partial }{\partial t}\right] \varphi (x,t)=0, 
\end{eqnarray} 
where ${n}=Na/l$. The normalization condition for the wave function is 
\begin{eqnarray}\label{n1}
\int_0^\infty dx |\varphi(x,t)|^2=1.
\end{eqnarray}

\subsection{Axially Symmetric Case}

The trap potential is given by  $  V({\bf r}) =\frac{1}{2}m \omega
^2(\rho^2+\lambda^2 z^2)$  where  $\omega$ is the angular frequency in the
radial direction $\rho$ and  $\lambda \omega$ that in  the axial direction $z$.
We are using the cylindrical coordinate system ${\bf r}\equiv (\rho,\theta,z)$
with $\theta$ the azimuthal  angle. In this case one can have quantized vortex
states with rotational motion around the $z$ axis \cite{11a}. In such a
vortex the atoms
flow with tangential velocity $\hbar L/(mr)$ such that each atom has quantized
angular momentum $\hbar L$ along $z$ axis.  The wave function can then be
written as: $ \Psi({\bf r};\tau)=\psi(r,z;\tau)\exp (iL\theta),$ with $L=0, \pm
1, \pm 2, ...$.
 
Using the above angular distribution of the wave function in (\ref{a}),  in
terms of dimensionless variables $x =\sqrt 2 r/l$,  $y=\sqrt 2 z/l$,   $t=\tau
\omega, $  $l\equiv \sqrt {\hbar/(m\omega)}$,  and   $   {
\varphi(x,y;t)} = x \sqrt{4\pi l^3/\sqrt 8}\psi(r,z;\tau), $ we get
\begin{eqnarray}\label{d}
 \biggr[& -& \frac{\partial^2}{\partial
x^2} + \frac{1}{x}\frac{\partial}{\partial x} -\frac{\partial^2}{\partial
y^2}+\frac{L^2}{x^2}
+\frac{1}{4}\left(x^2+\lambda^2 y^2-\frac{4}{x^2}\right) \nonumber \\
& + & 2\sqrt 2{n}\left|\frac {\varphi({x,y};t)}{x}\right|^2 
 -i\frac{\partial }{\partial t} \biggr]\varphi({ x,y};t) = 0.
\end{eqnarray}
The normalization condition  of the wave
function is \begin{equation}\label{5}  \int_0 ^\infty
dx \int _{0}^\infty dy|\varphi(x,y;t)|
^2 x^{-1}=1. \label{n2} \end{equation}

\subsection{Numerical Detail}
We solve the GP equation numerically employing a split-step time-iteration 
method using the Crank-Nicholson discretization scheme described
elsewhere \cite{murg}. We discretize the GP equation 
 spanning $x$ from 0 to 25 and $y$ from $-30$ to 30 in 
the axially symmetric case. In the spherically symmetric case the range of
$x$ integration was taken from 0 to 30.
This
was enough for having a negligible value of the wave function at the
boundaries. To calculate the solution for a specific nonlinearity $n$
the time iteration is started with the known harmonic 
oscillator solution of the GP equation for nonlinearity $n=0$. These
initial normalized solutions are 
\begin{equation} 
\varphi(x) = (2/\pi) ^{1/4} x e^{-x^2/4}
\end{equation} 
\begin{equation} 
\varphi(x,y) = \left[        \frac{2\lambda}{ \pi 2^ {2|L|} (|L|!)^2}
 \right] ^{1/4}{x^{1+|L|}}
e^{-(x^2+\lambda y^2)/4}
\end{equation} 
for  (\ref{c}) and (\ref{d}), respectively.
The nonlinearity 
is then slowly changed in steps of 0.0001 during each time iteration 
until the desired value of $n$ is attained. As a result  in the course of  
time
iteration
the initial $n=0$  solution 
slowly converges  towards the  final solution with the desired $n$. 

\subsection{Resonance Formation}

We study the formation of resonance in the oscillation of the
condensate as the scattering length is varied periodically. In the scaled
GP equations (\ref{c}) and (\ref{d}) the scattering length appears only
through the nonlinearity parameter $n=Na/l$. 
In particular
we consider the following variation of the nonlinearity parameter
\begin{equation}\label{e}
n= b+ c \sin (\Omega t),  
\end{equation}
where $b$ and $c$ are two constants and $\Omega$ is the frequency of
variation of the nonlinear parameter. 

In the GP equation
(\ref{c}), the scaled harmonic oscillator trap frequency is unity; the 
radial and axial trap frequencies of (\ref{d}) are 1 and $\lambda$,
respectively. For zero and small values of nonlinearity 
the natural frequency of oscillation of  (\ref{c}) is twice the scaled 
trap frequency, which is 2. The natural frequencies of
oscillation of
 (\ref{d}) in radial and transverse directions are  twice the scaled
trap frequencies, which are 2 and $2\lambda$, respectively. These natural
frequencies of oscillation have been verified experimentally \cite{7}, as
well as
in theoretical calculations \cite{ska}. The perturbation (\ref{e}) 
when applied to
the GP equation behaves like  a nonlinear external force acting in 
(\ref{c}) and  (\ref{d}). When the frequency $\Omega$ of the
perturbation coincides with the natural frequencies of oscillation of the
GP equation, resonant behaviors are expected. 

The classic example of resonance appears when the parameter $b$ 
in  (\ref{e}) is zero. In that case resonances are expected in the
spherically symmetric case for $\Omega =2$. In the axially symmetric case
they are expected for  $\Omega =2$ and  $\Omega =2\lambda$ in the radial
and axial oscillations, respectively. For $b\ne 0$, resonances also appear
provided that the constant $c$ is not very small. We study these
possibilities in the next sections.

\section{Result for Spherical  Trap}

First we consider the case with $b=0$ in  (\ref{e}). In this case the
maximum permitted value of $c$ is 0.575. For $c> 0.575$, the minimum of
nonlinearity $n$ could be less than $- 0.575$. It is known that no stable
solution of the spherically symmetric GP equation could be obtained for
$n< -0.575$ \cite{ex2,11,sk1}. The domain of values $n< -0.575$
corresponds
to attractive
interaction leading to collapse of the condensate. 

In this case 
the natural
frequency of oscillation of the condensate is 2 and resonances are
expected for values of $\Omega$ equaling multiples of 2. We shall mostly
concentrate here to the case $\Omega=2$. In figures 1 (a), (b) and (c) we
plot the root mean square  (rms) radii  $\langle x \rangle $ vs. time for
$c=0.1,
0.3$ and
$0.5$ and for $\Omega=2$.  We find that as $c$ increases the rms radii can
grow to a large value signaling a resonance. One interesting feature of
these resonances is that the oscillation of the rms radius can increase
and subsequently decrease and such a cycle  can repeat many times. A
larger
number of such growth and decay cycles can be accommodated in a given
interval of time as $c$ increases.

\begin{figure}[!ht]
\begin{center}
\includegraphics[width=0.5\linewidth]{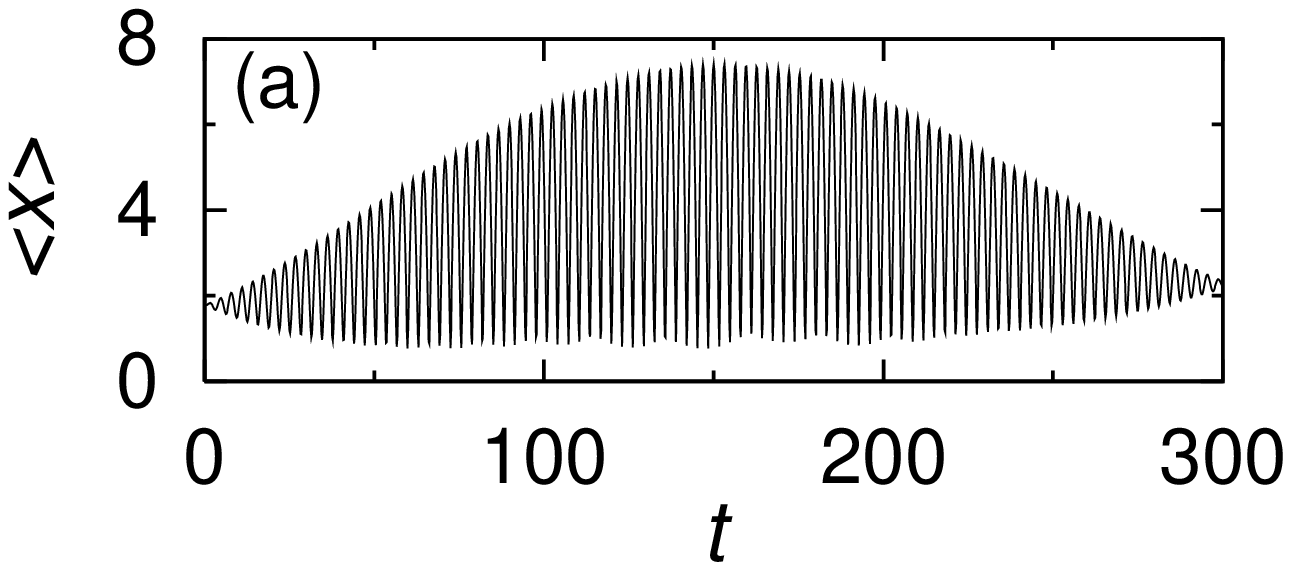}
\includegraphics[width=0.5\linewidth]{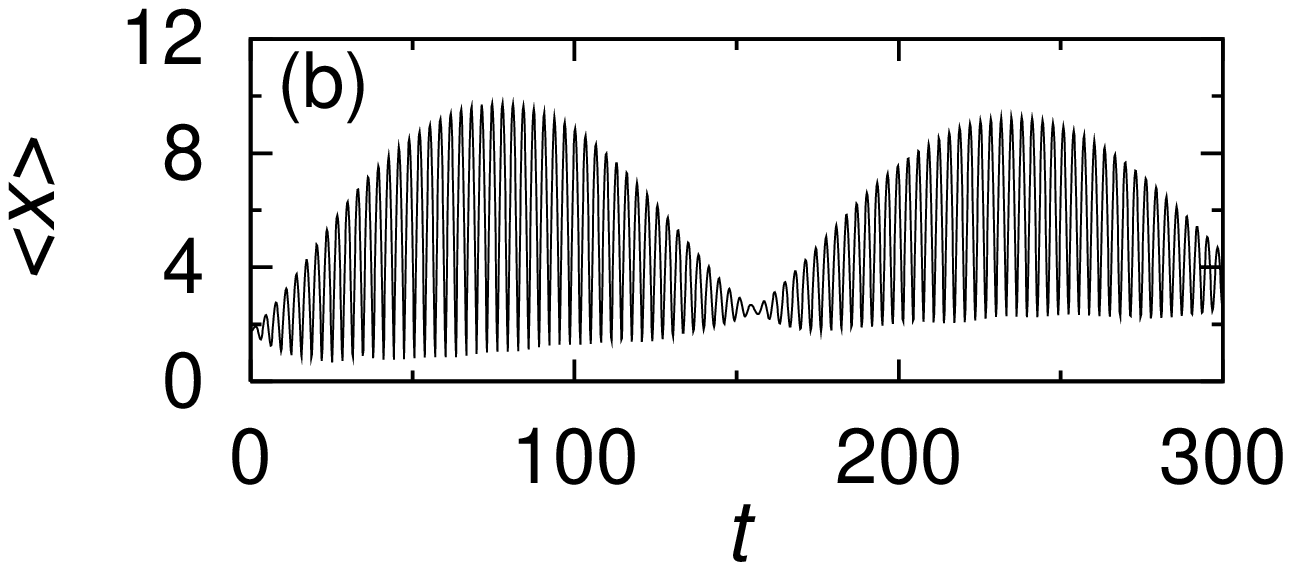}
\includegraphics[width=0.5\linewidth]{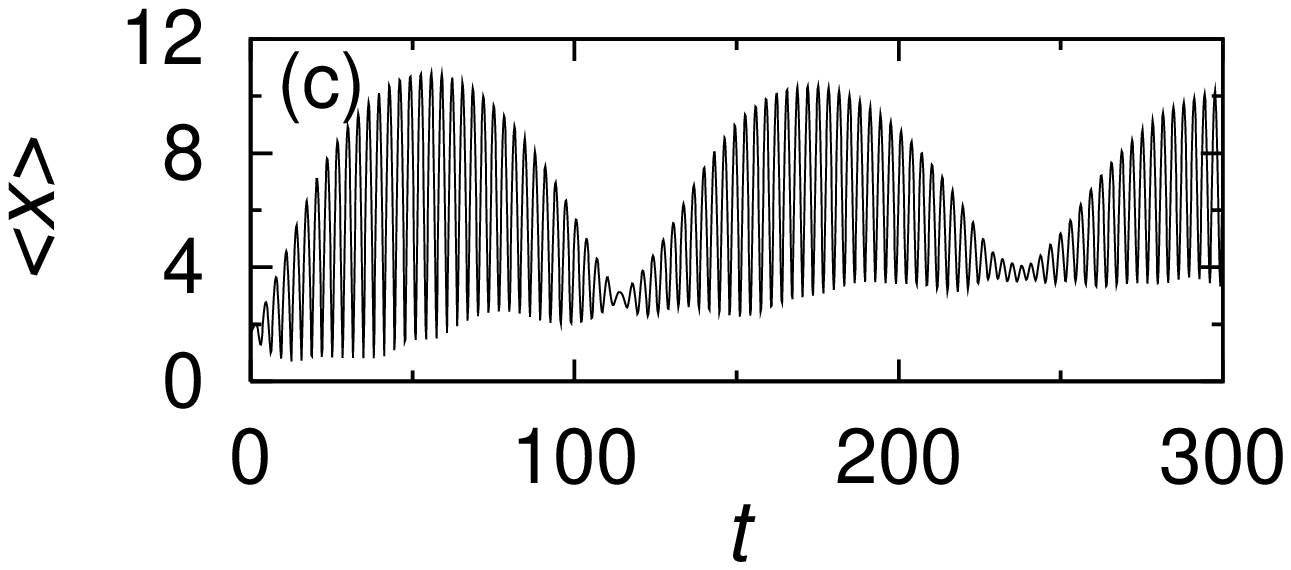}
\includegraphics[width=0.5\linewidth]{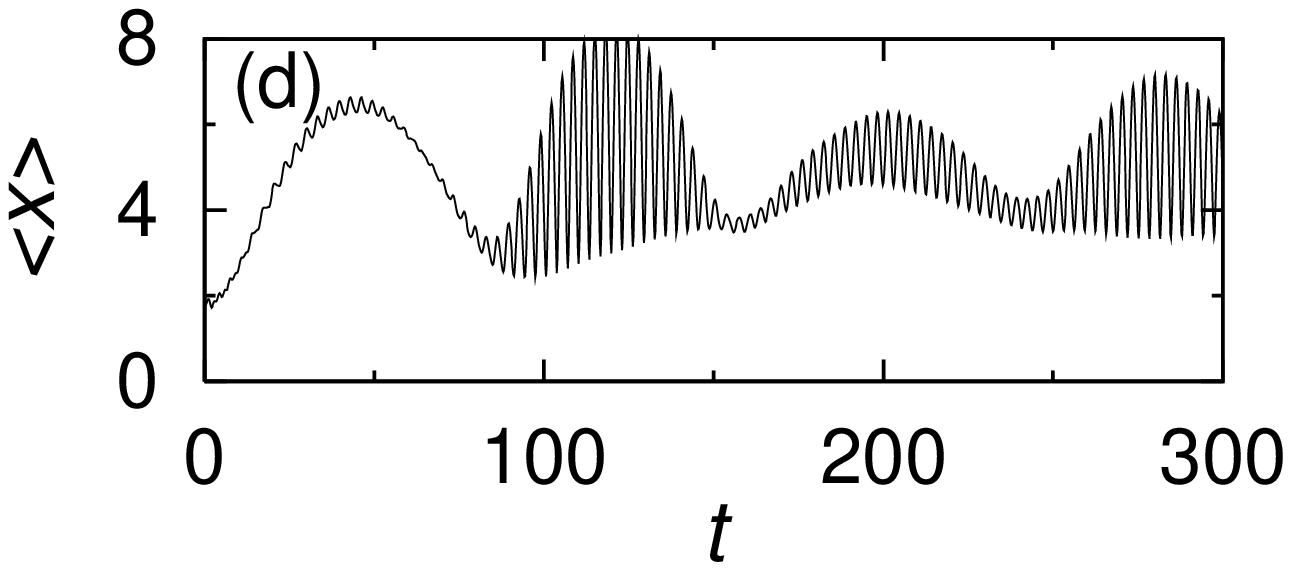}
\includegraphics[width=0.5\linewidth]{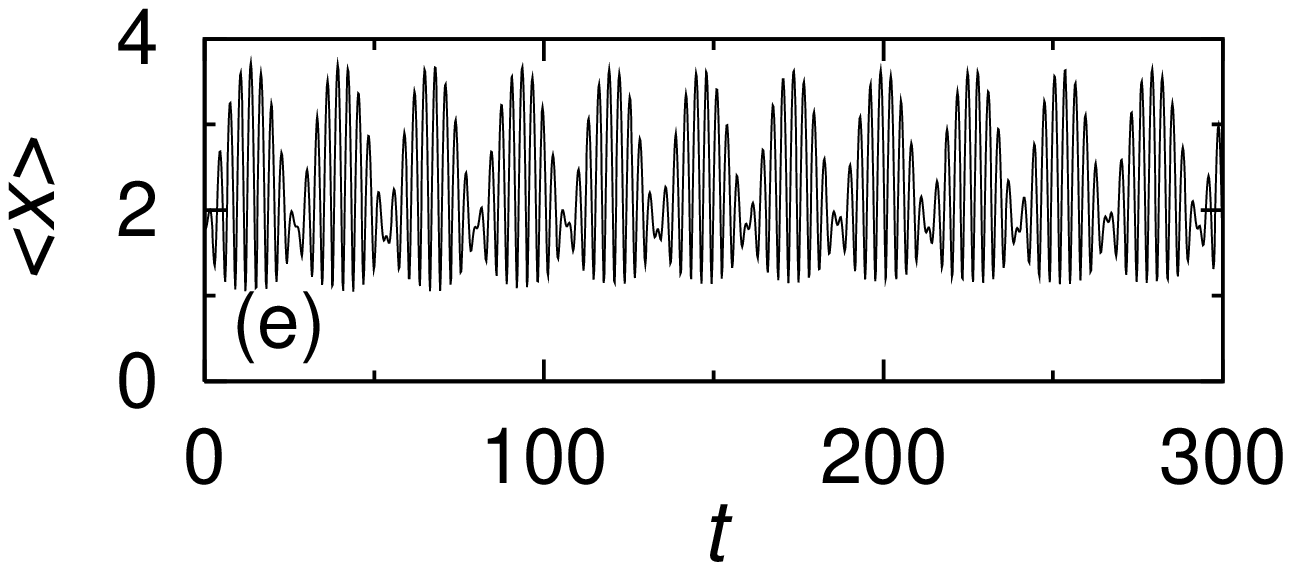}
\end{center}
 
\caption{The rms radii of the condensate $\langle x \rangle$ vs. time $t$
for $b=0$  and (a) $c=0.1$, $\Omega=2$, (b) $c=0.3$, $\Omega=2$, (c)
$c=0.5$, $\Omega=2$, (d) $c=0.5$, $\Omega=4$, (e) $c=0.5$, $\Omega=2.2$. }

\end{figure}
  
\begin{figure}[!ht]
\begin{center}
\includegraphics[width=0.5\linewidth]{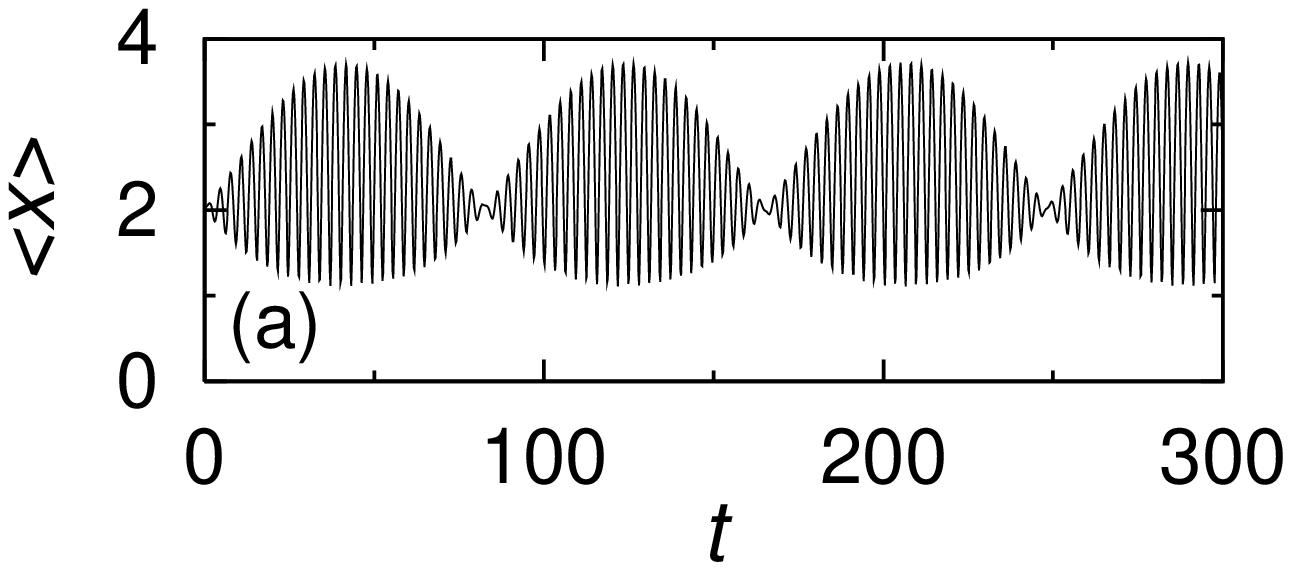}
\includegraphics[width=0.5\linewidth]{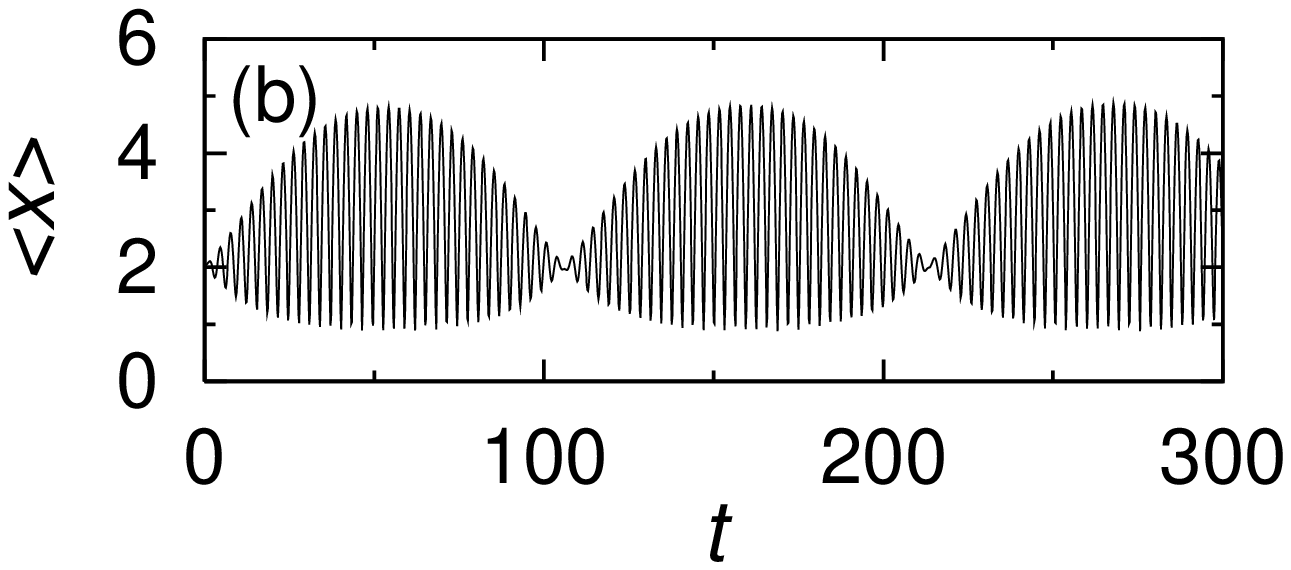}
\includegraphics[width=0.5\linewidth]{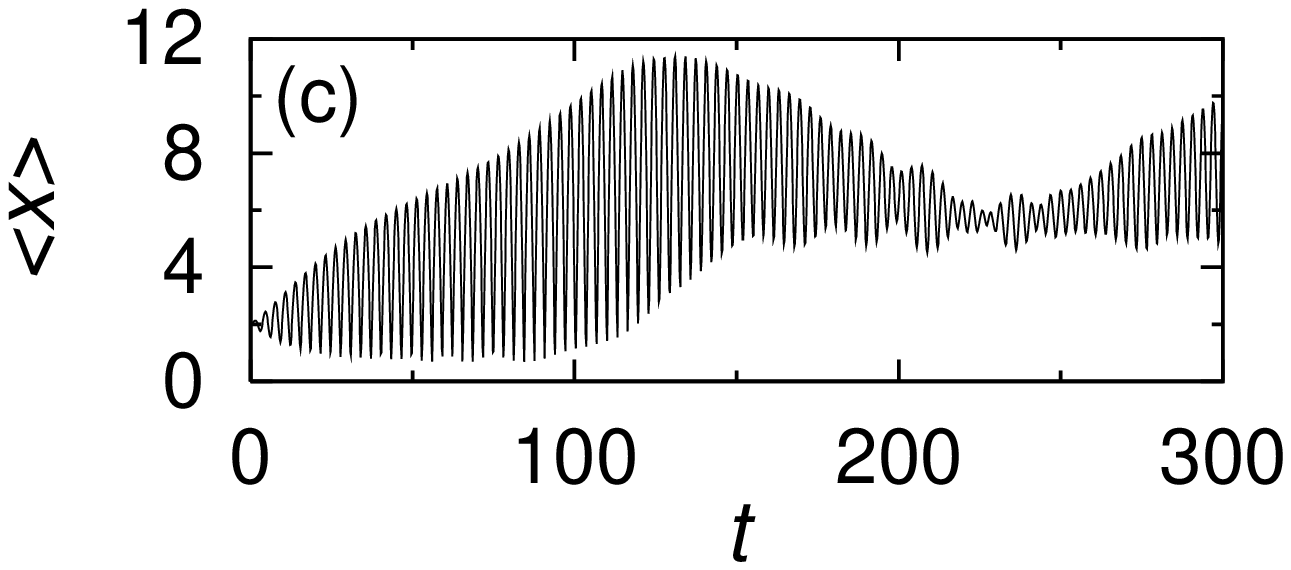}
\includegraphics[width=0.5\linewidth]{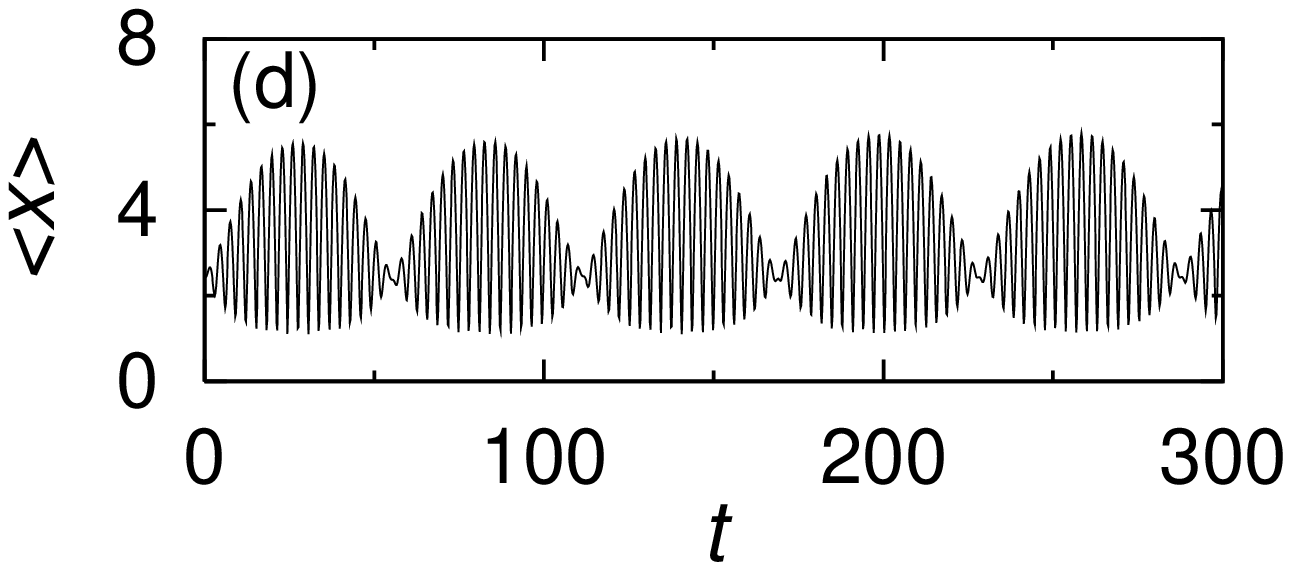}
\includegraphics[width=0.5\linewidth]{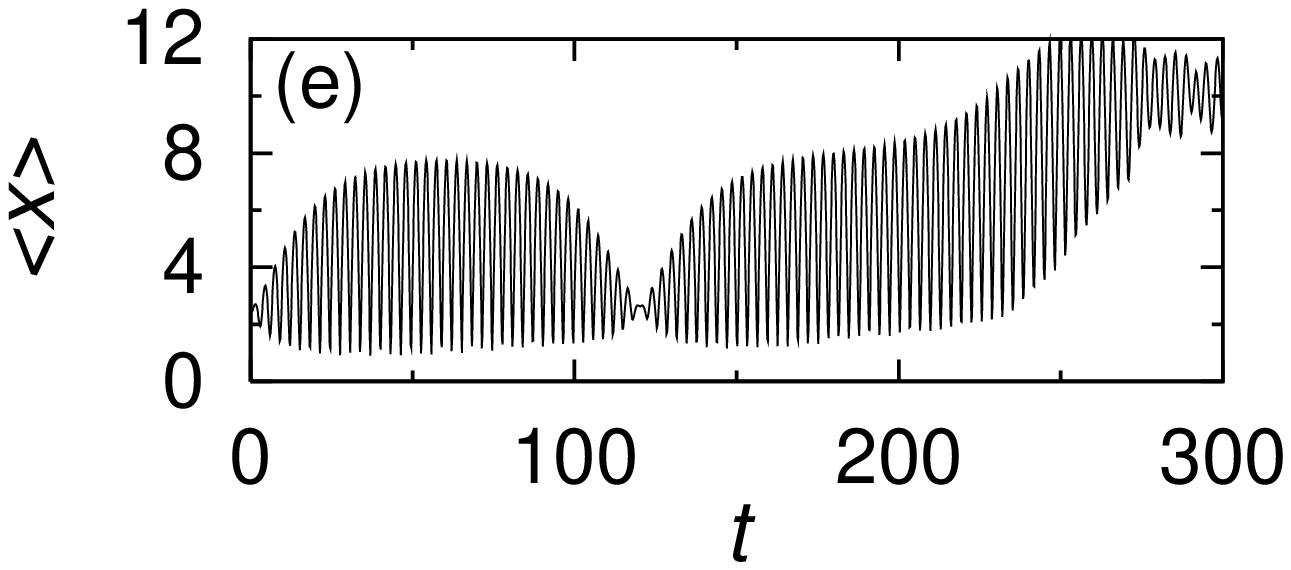}
\end{center}
 
\caption{The rms radii of the condensate $ \langle x \rangle $ vs. time
$t$ for $\Omega = 
2$  and (a) $b=1,c=0.3$ (b) $b=1,c=0.4$,
(c)
$b=1,c=0.5$,   (d) $b=5,c=2$, and (e)  $b=5,c=2.4$ } 

\end{figure}

In figure 1 (d) we plot the results for $c=0.5$ and $\Omega=4$. We find
that
again one can have large amplitudes of oscillation. Similar resonances
appear for $\Omega = 6, 8, ...$ etc. However, in this work we shall only
study the resonance of lowest order for $\Omega=2$. Finally, in figure 1
(e)
we present a off-resonance result for $c=0.5$ and $\Omega =2.2$. Compared
to figure 1 (c) with same values of $b$ and $c$, we find oscillation with
highly reduced amplitude in figure 1 (e) due to a variation of $\Omega$ to
an off-resonance value. If  $\Omega$ is taken further off-resonance the
oscillations disappear quickly.

In a problem of resonance of an uncoupled oscillator driven by a periodic 
external force the oscillator oscillates in phase with the external
force. Hence it gains energy  during every cycle of oscillation and the
amplitude of oscillation keeps on increasing. So a pertinent question to
ask is how can the amplitude of oscillation decrease in the present
problem. In this nonlinear problem the phase between the  oscillation at
resonance and the external force varies with time. Hence during a certain
interval of time the two are in phase and the amplitude of oscillation
grows. After the amplitude attains a certain value, the external force and
the oscillation become out of phase and the  system loses energy in each
cycle. Eventually, the 
amplitude of oscillation
reduces to a small value and this growth and decay cycle may repeat  for
a long time.  Due to the nonlinearity in the equation the system moves
in  and  out  of resonance when a constant frequency drive is
applied. This
leads  to  oscillations  of  varying  time-dependent  amplitude  in the
rms
widths.

Next we study the resonance in  the situation with $b\ne 0$ and $\Omega
=2$.  For a nonzero $b$, pronounced resonances appear as $c$ is increased
from 0. In figures 2 (a), (b), and (c) we plot $\langle x \rangle $
vs. $t$
for $b=1$, and 
$c= 0.3, 0.4$ and 0.5, respectively. The amplitude of oscillation
increases
as $c$ is increased. We again see the growth and decay cycles for small
values of $c$. However, such periodic  growth and decay cycles are lost
with the increase in $c$ as one can find in figure 2 (c) for $c=0.5$.   

In figures 2 (d) and (e) we show the results for $b=5$ and $c=2$ and 2.4,
respectively. With the increase of $c$, the amplitude of oscillation has
increased from figure  2 (d) to  2  (e). However, the periodic  
growth and decay cycles have been  lost in this transition.  

There are two time scales in each of the
figures. The first one is the driving frequency and the second one is
related to the nonlinear term in the GP equation. For the linear
problem only the first time
scale appears  and the second frequency is absent. At resonance the second
frequency should 
increase with the effective nonlinearity of the system although a simple 
relation between the two cannot be obtained.  For $b=0$ the effective
nonlinearity increases with increasing $c$ and at resonance the second
frequency is
expected to
increase as $c$ increases. However, for a positive $b$ the effective
nonlinearity is large when $c$ is large and at resonance  second frequency
is
expected to decrease as  $c$ increases.  
These  interesting features of the growth and decay cycles emerges from
figures
1
and 2. 
At resonance for $b=0$ 
more growth and decay cycles appear for
large values of $c$. 
However, quite expectedly for $b\ne 0$  more growth and decay cycles
appear for  small values of $c$.

\section{Result for Axial Trap}

Next we continue our discussion of resonance to axially symmetric
traps.  We find that when the external
frequency $\Omega$ coincides with the natural frequency of radial
($\omega =2$) or axial ($\omega = 2 \lambda$) oscillation, there is
resonance in  radial or axial oscillation, respectively. 
The general nature of
this oscillation is similar to that in the spherically symmetric case.
We studied the
resonance dynamics for different sets of parameters. 
However, in the following we present results only for $b=0$ and $c=0.5$
for angular momenta $L= 0$ and 1.  We consider both pancake
($\lambda > 1$)
and cigar-type ($\lambda < 1$) deformation of the condensate in our
study. We recall that $\lambda =1$ corresponds to the spherical
 case considered in the last section.

\subsection{States with zero angular momentum}

\begin{figure}[!ht]
\begin{center}
\includegraphics[width=0.5\linewidth]{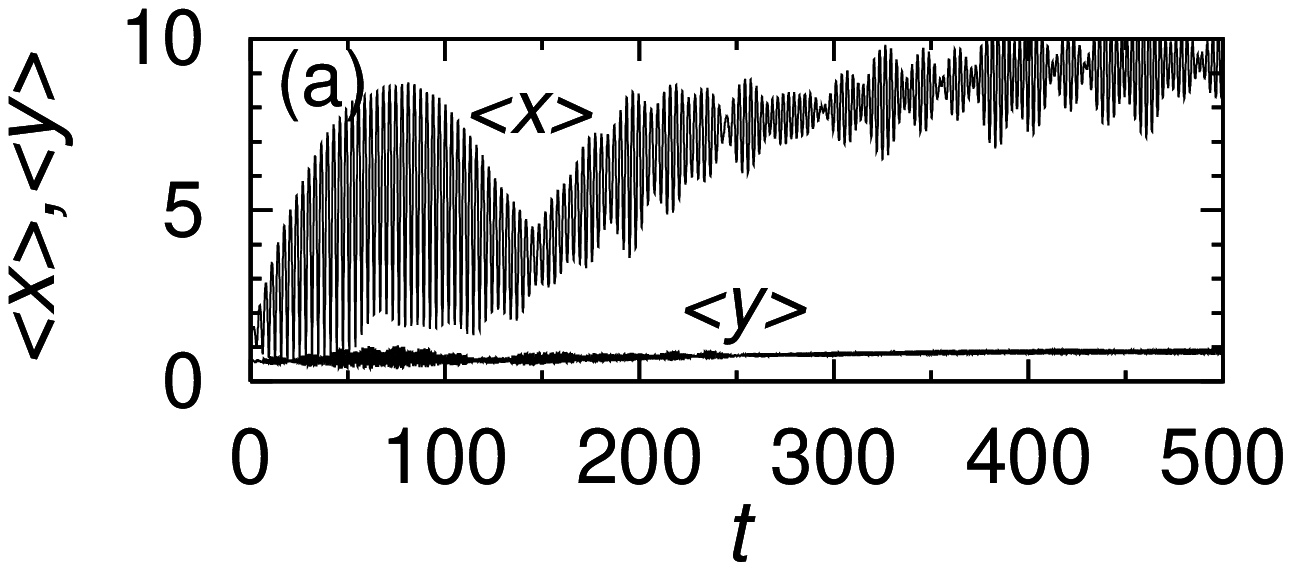}
\includegraphics[width=0.5\linewidth]{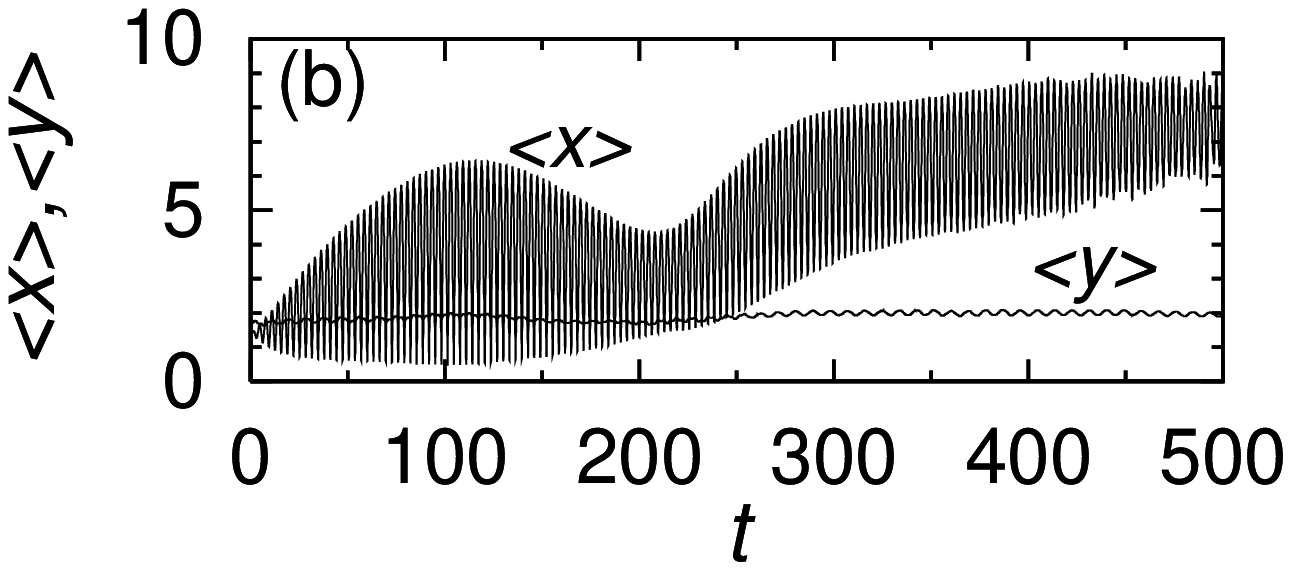}
\includegraphics[width=0.5\linewidth]{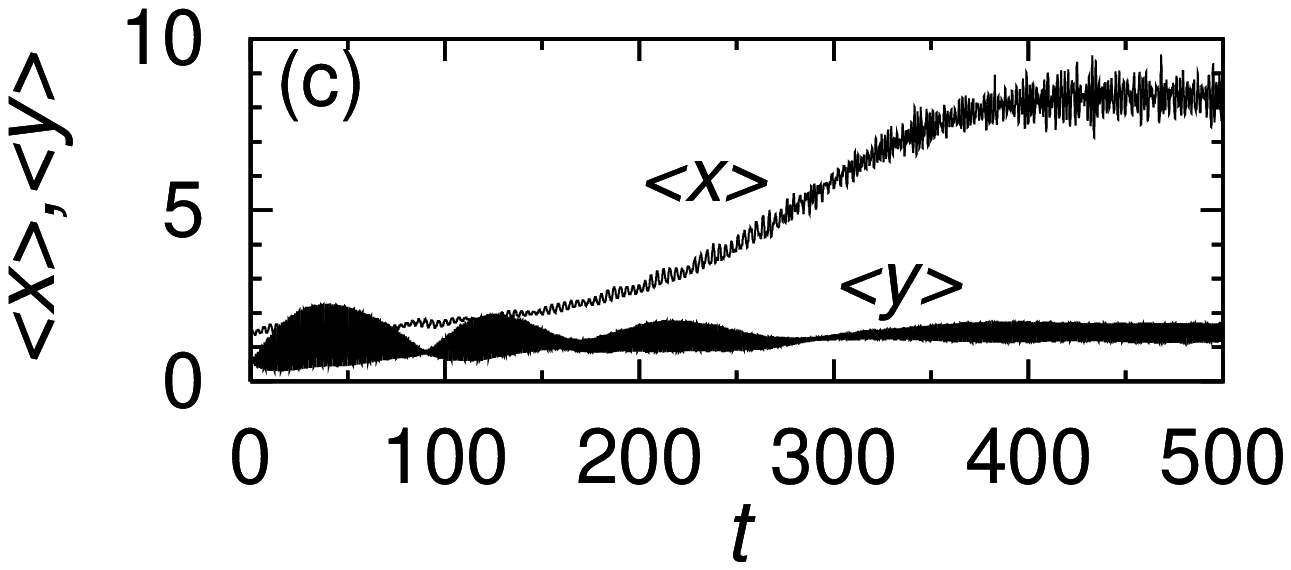}
\includegraphics[width=0.5\linewidth]{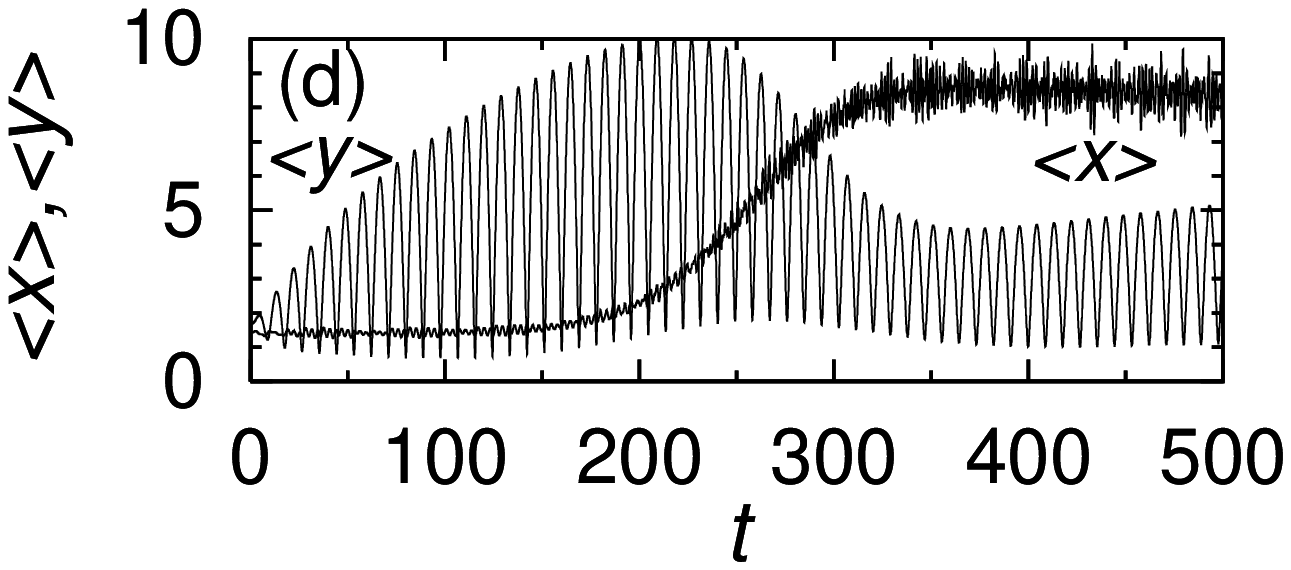}
\end{center}
 
\caption{The radial and axial rms sizes  of the condensate $ \langle x
\rangle $
and $ \langle y \rangle $, respectively,  vs. time $t$ for $b=0 ,$
$c=0.3$, 
$L=0$ and for 
(a) $\lambda = \sqrt{8}$, $\Omega=2$, 
(b) $\lambda = 1/\sqrt{8}$, $\Omega=2$, 
(c) $\lambda = \sqrt{8}$, $\Omega=2\lambda$, 
(d) $\lambda = 1/\sqrt{8}$, $\Omega=2\lambda $. }

\end{figure}

Here,  for angular momentum $L=0$, 
we consider the following two values of the axial parameter $\lambda =
\sqrt{8}$  
and $1/\sqrt{8}$ as some of the early experiments \cite{1} were performed
with
these  anisotropies.  Also, these values of $\lambda$ correspond to
substantial contraction and elongation along the axial direction and
hence to a substantial deviation from the spherical symmetry. We
present results for $\Omega = 2$ and $2\lambda$ in each of these cases. 

In figures  3 (a), (b), (c), and (d) we plot the rms radial and axial
sizes
$ \langle x \rangle $ and $ \langle y \rangle $
for  
$\lambda = \sqrt{8}$ and $1/\sqrt{8}$ and for $\Omega =2$ and 
$2\lambda$.  
In figures 3 (a) and (b) resonance in radial oscillation 
is achieved for $\Omega =2$. This results in the large and rapidly varying 
rms radii $ \langle x \rangle $ at resonance and a small and slowly
varying rms size 
$ \langle y  \rangle$ off the resonance.  In figures 3 (a) and (b) there
is
one complete
growth and
decay cycle for $ \langle x  \rangle$. Then the resonance is manifested by
a
different 
type of oscillation in $ \langle x  \rangle$. Similar behavior was also
found in
the 
spherically symmetric case, e.g., in figure 2 (e). 

In figures 3 (c) and (d) resonance in axial oscillation
is achieved for $\Omega =2\lambda$.
In these cases the axial size $ \langle y  \rangle$ executes resonance
oscillation, whereas the rms radius $  \langle x  \rangle$
increases  
in size representing a swelling.
In figure 3 (c) the axial trapping frequency $\lambda 
= \sqrt{8}$ is large, which  makes the  axial oscillation  less likely
than
in the case presented in figure 3 (d), where the small  axial trapping
frequency $\lambda  =1/  \sqrt{8}$ makes the axial oscillation more
favorable.  Consequently, the amplitude at the  resonance in axial
oscillation in figure 3 (d)  is about four times more than that in  figure 
3
(c). 

From figures 3 (a) and (b) we find that at the resonance of radial
oscillation the axial rms size $  \langle y \rangle $ remains quite
constant.  
We find  in figures  3 (c) and (d) that at the resonance of axial
oscillation the radial rms radius $ \langle x  \rangle $ varies more with
time.  
One interesting feature has appeared in figures 3 (c) and (d), where for 
$\lambda = 1/\sqrt{8}$  the radial rms radii $\langle x \rangle $ vary
with time 
corresponding to a slow radial swelling accompanied by small oscillation 
and not an  oscillation with
large amplitude.
We verified, as in the spherically symmetric case, that a smaller value of
the
constant $c$ leads to many growth and decay cycles of the amplitude at
resonance in the axially symmetric case (however, not shown explicitly
here). 

\subsection{Vortex States}

Here for angular momentum $L=1$ we consider the effect of a periodic
variation of the scattering
length via  (\ref{e}) on the quantized vortex states. 
Such states are
of great
importance to the study of the BEC, as they are intrinsically related to
the existence of superfluidity \cite{xx}. The experimental detection of
vortex
states in BEC \cite{ex3} makes the resonant oscillation of the vortex
states under a
periodic variation of scattering length experimentally observable.

\begin{figure}[!ht]
\begin{center}
\includegraphics[width=0.5\linewidth]{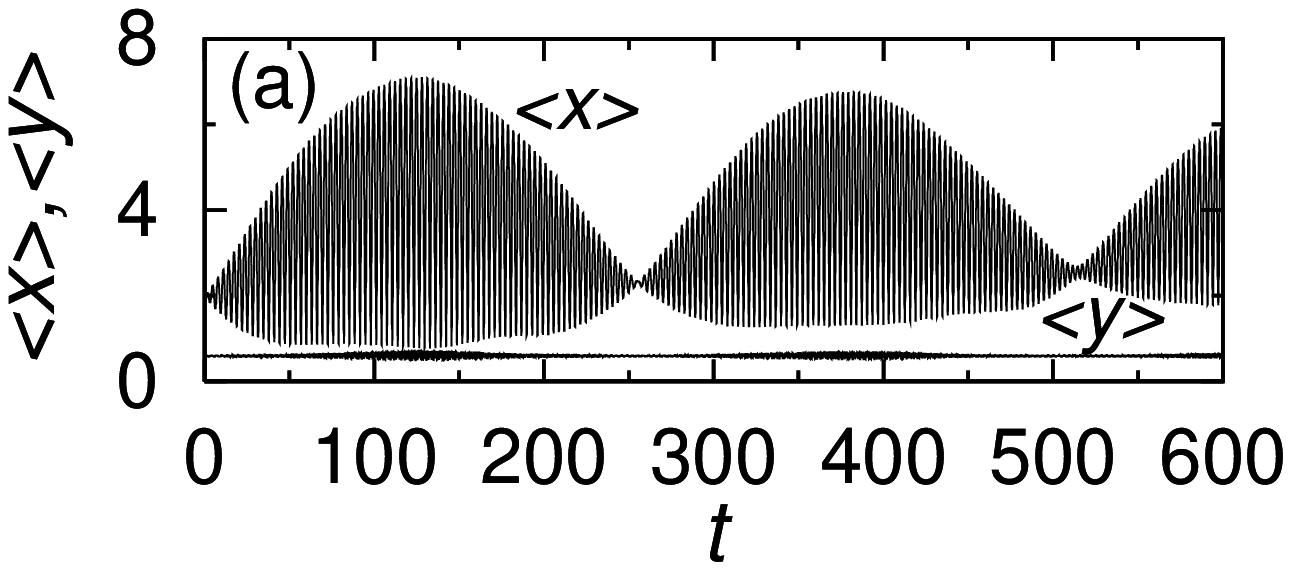}
\includegraphics[width=0.5\linewidth]{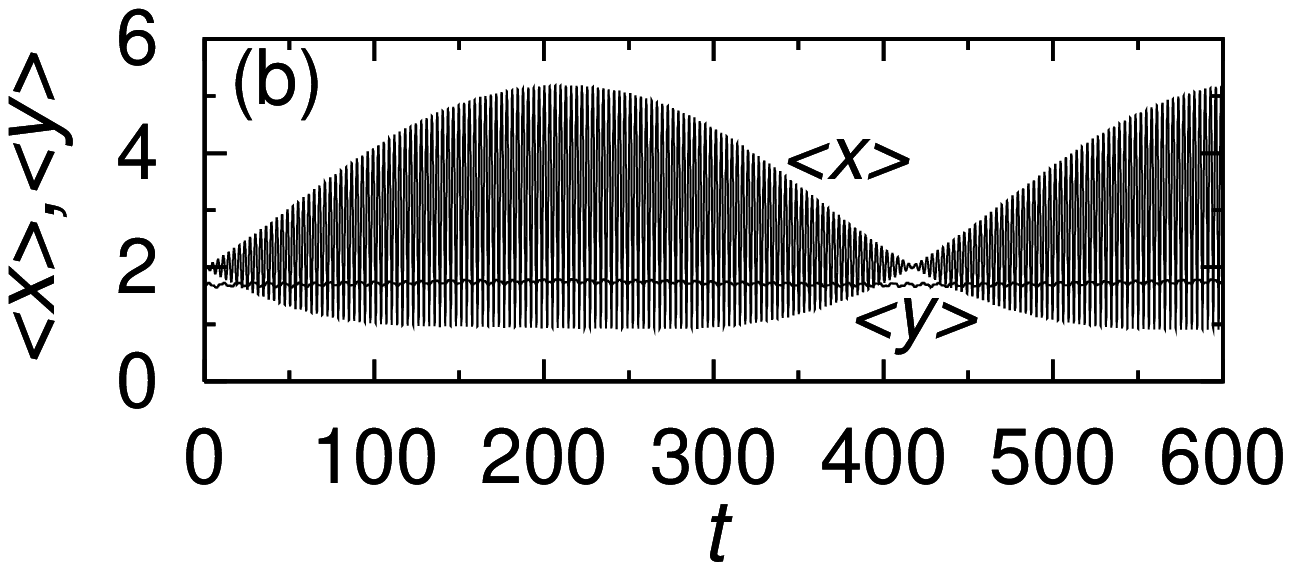}
\includegraphics[width=0.5\linewidth]{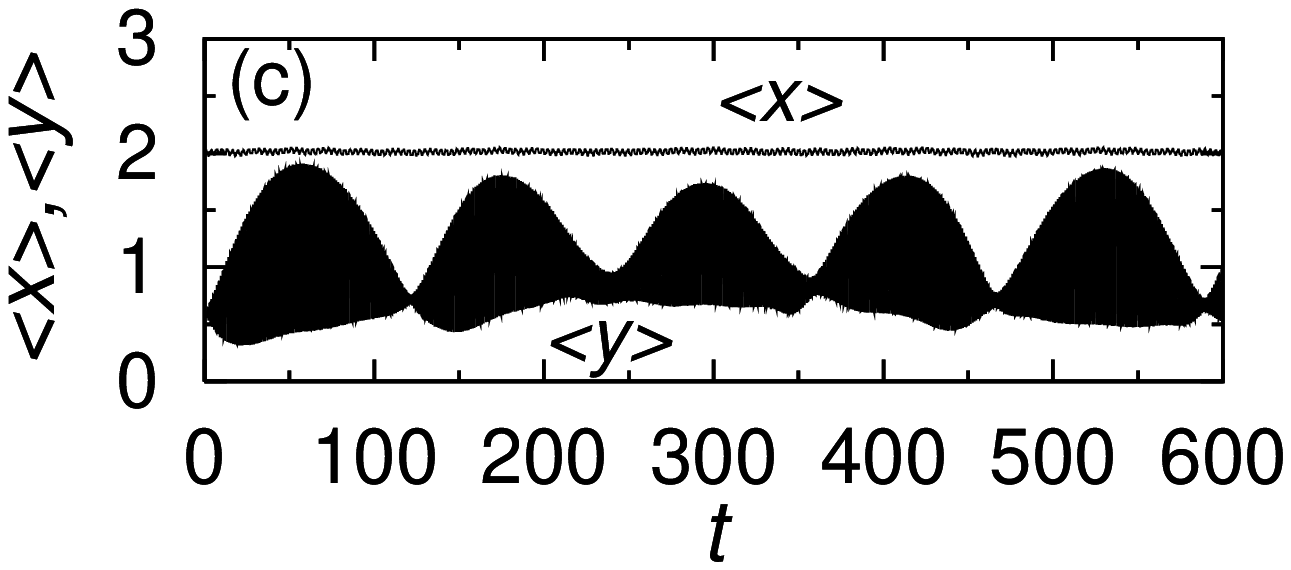}
\includegraphics[width=0.5\linewidth]{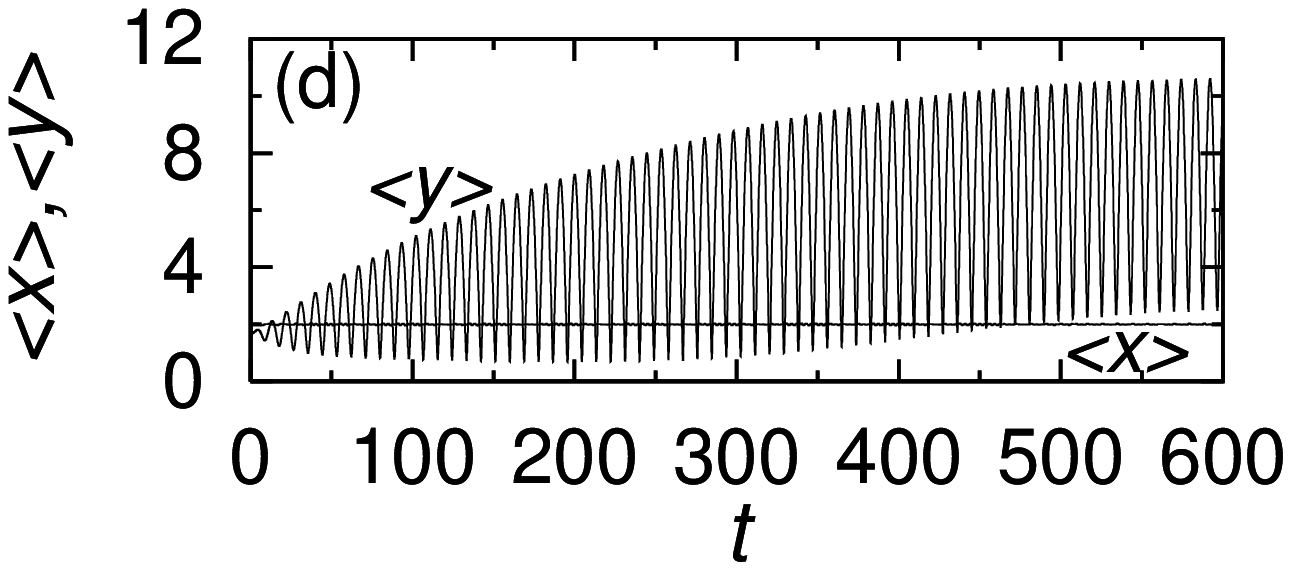}
\end{center}
 
\caption{The radial and axial rms radii of the condensate $  \langle x
\rangle $
and $  \langle y  \rangle $, respectively,  vs. time $t$ for $b=0 ,$
$c=0.3$, $L=1$  and for 
(a) $\lambda = \sqrt{8}$, $\Omega=2$, 
(b) $\lambda = 1/\sqrt{8}$, $\Omega=2$, 
(c) $\lambda = \sqrt{8}$, $\Omega=2\lambda$, 
(d) $\lambda = 1/\sqrt{8}$, $\Omega=2\lambda $. 
}

\end{figure}

To illustrate the resonance we consider the lowest angular excitation of
the BEC rotating round the axial direction with each atom possessing
angular momentum $\hbar$ corresponding to $L=1$. Again we consider the
anisotropy parameters $\lambda = \sqrt{8} $ and $1/\sqrt{8}$ and the
external frequencies $\Omega =2$ and $2\lambda$.  In figures 4 (a), (b),
(c), and (d) we plot the rms radial and axial sizes $ \langle x  \rangle $
and $  \langle y  \rangle $ in
these cases. These figures should be compared with the respective zero
angular momentum cases shown in figures 3 (a), (b), (c), and (d). 

In the case of vortex states very orderly (periodic) growth and decay
cycles of the resonating amplitudes appear for $b=0$ and $c=0.3$ in
 (\ref{e}). The periodic  growth and decay
cycles of the resonating amplitudes are replaced by irregular ones as in 
figures 3 (a), (b), (c), and (d) as the value of the constant $c$ is
increased (not shown explicitly in this paper). 
The off-the-resonance 
amplitudes $-$  $\langle y \rangle$ in figures 4 (a) and (b) and  $\langle
x
\rangle$  in figures 4 (c) and (d) 
$-$
remain constant without variation in all cases. However, the
period of growth and decay
cycles of the resonating amplitude varies from one case to another. In all
four cases we have plotted the amplitudes up to 600 dimensionless time 
units. This accommodates about 2.3  growth and decay
cycles of the resonating amplitude in figure 4 (a), 1.5 cycles in figure 4
(b), 5 cycles in figure 4 (c), and 0.5 cycles in figure 4 (d). 

Comparing the situation of figure  4 (c)  with that in figure 4 (d) we see 
that there is a reduction in the axial trapping frequency as we pass from 
figure  4 (c) to  figure 4 (d). This reduction in the axial trapping
frequency 
should favor the axial oscillation at resonance  in
figure 4 (d).  Consequently, we find that the amplitude of axial
oscillation 
at resonance is much larger in  figure 4 (d) compared to that in figure  4
(c). Similar effect was also observed in figures 3 (c) and (d) in the 
$L=0$ case.

\section{Summary}

We have studied the occurrence of resonant oscillation in a BEC
due to a periodic variation in the atomic scattering length $a$. 
A variation in the atomic scattering length near a Feshbach resonance
can be and has been achieved experimentally \cite{fesh} by varying an
external 
magnetic field. This has been executed experimentally in producing a
sudden jump in the scattering length from positive (repulsive) to
negative (attractive) which has led to collapse and  explosion in the
condensate \cite{7}
and several novel phenomena \cite{8}.  Here we demonstrate that
interesting
resonating oscillation can be generated in the condensate if the
scattering length is varied periodically with a given  frequency. 
This oscillation may be verified experimentally and novel and interesting 
phenomena may emerge as a consequence of  such verification. 

Mathematically, the periodically oscillating scattering length in the GP
equation can 
be considered as an external nonlinear periodic driving force in the 
linear Schr\"odinger equation, which is an interesting problem of quantum 
mechanics and deserves special attention. Here we have considered  a  full 
numerical approach to its solution and have revealed interesting results.
One could also employ an analytic perturbative procedure for its solution
when the coefficient of the external force is small. This would be a
problem of future interest.

We found that 
resonances with novel feature can appear in the oscillation of the BEC
from a periodic 
variation of the scattering length when the frequency of the oscillation 
of scattering length $\Omega$ coincides with a multiple of the natural
frequency of oscillation of
the harmonically trapped BEC. For a spherically symmetric condensate, 
the  natural
frequency of oscillation of the BEC is twice the trapping frequency
($=2\omega$).  Hence resonances in the BEC oscillation appear for 
$\Omega = 2\omega, 4\omega, 6\omega,...$ etc. For an axially symmetric
BEC, the  natural
frequency of oscillation of the BEC in radial and axial directions are
twice the trapping frequencies in these directions, which are $=2\omega$
and $=2\lambda \omega$, respectively. Consequently, resonances in the BEC
oscillation in the radial and axial directions appear for
$\Omega = 2\omega, 4\omega, 6\omega,...$ and 
$\Omega = 2\lambda\omega, 4\lambda \omega, 6\lambda \omega,...$,
respectively. The interesting feature of these resonances is that even in
the absense of any damping (dissipative force), the amplitude of
oscillation at resonance can grow and reduce with time. Such growth and
decay cycle  of the amplitude may  repeat  several times. The period of
such growth and decay cycles is $\sim 100$ units of dimensionless time,
which under usual experimental condition with $\omega \simeq 500$ Hz/s
corresponds to about  0.2 s $-$ a period that can be precisely measured
and compared with theoretical prediction to test the mean-field theory. 
We hope that the
present investigation may stimulate further theoretical and experimental
studies on this topic.

\ack

The work is supported in part by the Conselho Nacional de Desenvolvimento
Cient\'\i fico e Tecnol\'ogico and Funda\c c\~ao de Amparo \`a Pesquisa do
Estado de S\~ao Paulo of Brazil.


\section*{References}
\begin{thebibliography}{10} 
\bibitem{1}  

  Anderson M H,  Ensher J R,  Matthews M R, 
Wieman C E and  Cornell E A 1995 {\it Science} {\bf 269} 198 

 
Ensher J R,   Jin D S,  Matthews M R,  Wieman C E and
  Cornell E A 1996 {\it
Phys.
Rev. Lett.  } {\bf 77} 4984 

   Davis K B,  Mewes M O, 
Andrews M R,  van Druten N J,  Durfee D S,  Kurn D M and
 Ketterle W  1995
{\it Phys.
Rev. Lett.}  {\bf 75} 3969 

  Fried D G,  Killian T C,
Willmann L,  Landhuis D,   Moss S C,  Kleppner D, 
 Greytak T J 1998 {\it
Phys.
Rev. Lett.}   {\bf 81} 3811 

 
\bibitem{ex2} Gerton J M,  Strekalov D,  Prodan I and  Hulet R G 2001 {\it
Nature} {\bf 408} 692 


 Bradley C C,  Sackett C A,
  Tollett J J  and  Hulet R G 1995 {\it
Phys. Rev. Lett. }  {\bf 75} 1687 



 Bradley C C,  Sackett C A,
 and  Hulet R G 1997 {\it
Phys. Rev. Lett. }
   {\bf 78} 985 


 
\bibitem{11}  Dalfovo F,  Giorgini S,  Pitaevskii L P and
 Stringari S 1999 {\it
Rev. Mod.  Phys.}  {\bf 71} 463 

\bibitem{11a} 
Adhikari S K 2002 {\it Phys. Rev. A} {\bf 65}  033616 

\bibitem{11b}  Kagan Yu,  Muryshev A E and
 Shlyapnikov G V 1998 {\it Phys. Rev. Lett.} {\bf 81} 933

 Saito H and
 Ueda M  2001 {\it Phys. Rev. Lett.} {\bf 86} 1406 

 Esry B D,
 Greene C H
and
 Burke, Jr. J P  1999 {\it Phys. Rev. Lett.} {\bf 83} 1751 
 
 Ruprecht P A,
 Holland M J,  Burnett K  and  Edwards M 1995 {\it
Phys. Rev. A}  {\bf 51} 4704 

 Eleftheriou A and Huang K 2000
 {\it Phys. Rev. A} {\bf 61} 043601 

  Dalfovo F and Stringari S 1996 {\it
Phys. Rev. A} {\bf 53} 2477 

  Houbiers M and  Stoof H T C
1996 {\it Phys. Rev. A}
{\bf 54} 5055

   Holland M and
 Cooper J 1996   {\it Phys. Rev. A} {\bf 53}  R1954 

 
\bibitem{2} Gross E P 1961 {\it  Nuovo Cimento} {\bf 20} 454


  
Pitaevskii L P  1961 {\it Zh. Eksp. Teor. Fiz.} {\bf 40} 646
 [1961 {\it Sov.  Phys.  JETP}
{\bf 13} 451]


 Leggett A J 2001
{\it Rev. Mod. Phys.} {\bf 73} 307 


\bibitem{5} Jin D S,  Ensher J R,  Matthews M R,  Wieman C E and
 Cornell E A 1996 {\it Phys. Rev. Lett.} {\bf 77} 420 

 Jin D S,  Matthews M R,   Ensher J R,   Wieman C E and 
 Cornell E A  1997 {\it Phys. Rev. Lett.}
  {\bf 78} 764 

\bibitem{6} Garc\'ia-Ripoll J J,   P\'erez-Garc\'ia V M and
 Torres P 1999 {\it
Phys. Rev. Lett.} {\bf 83} 1715 

 
Yukalov V I, Yukalova E P and Bagnato V S 2002 \PR A {\bf 66} 043602

\bibitem{7} Donley E A,  Claussen N R,  Cornish S L,  Roberts J L,
 Cornell E A and  Wieman C E 2001 {\it Nature (London)} {\bf 412} 295

 Roberts J L,  Claussen N R,  Cornish S L,  Donley E A,
 Cornell E A
and  Wieman C E 2001 {\it Phys. Rev. Lett.} {\bf 86} 4211 

 Claussen N R,  Donley E A,  Thompson S T and  Wieman C E  2002  {\it
Phys. Rev. Lett.} {\bf
89}
010401 

\bibitem{8}
 Strecker K E,  Partridge G B,  Truscott A G and
 Hulet R G  2002 {\it Nature (London)}  {\bf 417} 150


  Khaykovich L, 
Schreck F,  Ferrari G,  Bourdel T,  Cubizolles J,  Carr L D,
 Castin Y and 
 Salomon C  2002 {\it Science} {\bf 296} 1290 


\bibitem{9}  Adhikari S K 2002 {\it Phys. Lett. A}  {\bf 296} 145


 Saito H
and  Ueda M 2002  {\it   Phys. Rev. A}
{\bf 65} 033624 

 Santos L and
 Shlyapnikov G V 2002 {\it  Phys. Rev. A} {\bf 66} 011602

 Duine R A and
 Stoof H T C  2001 {\it Phys. Rev. Lett.} {\bf 86} 2204 

\bibitem{9a}  
 Carr L D
and  Castin Y 2002   {\it Phys. Rev. A}   {\bf 66}  063602  

 Leung V Y F,  Truscott A G and  Baldwin  K G H  2002   {\it Phys. Rev. A}
{\bf 66}  061602


 Salasnich L,  Parola A and Reatto L 2002   {\it Phys. Rev. A}
{\bf 66}  043603 

 Al Khawaja U,  Stoof H T C,  Hulet R G,
 Strecker K E and  Partridge G B 2002    {\it Phys. Rev. Lett.} 
{\bf 89} 200404


Adhikari S K  2002   e-print cond-mat/0207171



\bibitem{symon}  Symon K R 1953  {\it Mechanics},
(Reading: Addison-Wesley), Ch. 4.

\bibitem{fesh}  Inouye S,  Andrews M R,  Stenger J,  Miesner H J,
 Stamper-Kurn D M and  Ketterle W 1998 {\it  Nature (London)} {\bf 392}
151

 Stenger J,  Inouye S,   Andrews M R,   Miesner H J,
 Stamper-Kurn D M  and
 Ketterle W  1999 {\it Phys. Rev. Lett.} {\bf 82} 2422 



\bibitem{abd}  Abdullaev F K,  Bronski J C and  Galimzyanov R M 2002
 e-print cond-mat/0205464.




\bibitem{ska} 
Adhikari S K 2002 {\it Phys. Rev. A} {\bf 66} 043601

Adhikari S K 2002 {\it Phys. Rev. A} {\bf 66} 013611


\bibitem{murg} 
Adhikari S K and  Muruganandam P  2002 {\it
J. Phys. B: At. Mol. Opt. Phys.} {\bf 35}
2831 

Adhikari S K and  Muruganandam P  2003 {\it
J. Phys. B: At. Mol. Opt. Phys.} {\bf 36}
409 


\bibitem{sk1} Adhikari S K 2002 {\it  Phys. Rev. E} {\bf 65} 016703 

 Dodd R J,  Edwards M,  Williams C J,  Clark C W,  Holland M J,
 Ruprecht P A and  Burnett K 1996 {\it  Phys. Rev. A} {\bf 54} 661 


\bibitem{xx}Feder D L,  Clark C W and  Schneider B I 1999
{\it Phys. Rev. Lett.} {\bf 82} 4956 

  Dum R,  Cirac J I,
 Lewenstein M and  Zoller P 1998 {\it
Phys. Rev. Lett.}   {\bf 80} 2972


Feder D L and   Clark C W  2001
{\it Phys. Rev. Lett.}   {\bf 87} 190401


\bibitem{ex3} Madison K W,  Chevy F,  Wohlleben W and  Dalibard J 2000
{\it 
Phys. Rev. Lett.} {\bf 84} 806 

 Matthews M R, 
Anderson B P,  Haljan P C,  Hall D S,   Wieman C E and
 Cornell E A  1999 {\it
Phys. Rev. Lett.}
 {\bf 83} 2498 

 Abo-Shaeer J R, Raman C, 
 Vogels J M and  Ketterle W 2001 {\it Science} {\bf 292} 476 





 
 
 






\end{thebibliography}
\end{document}